\documentstyle[preprint,revtex]{aps}
\begin{document}
\draft
\preprint{LSUHE No. 142-1993 }
\def\overlay#1#2{\setbox0=\hbox{#1}\setbox1=\hbox to \wd0{\hss #2\hss}#1%
\hskip -2\wd0\copy1}
\begin{title}
SU(2) finite temperature phase transition \\
and Michael sum rules
\end{title}
\author{ Yingcai Peng and Richard W. Haymaker}
\begin{instit}
Department of Physics and Astronomy, \\
Louisiana State University, Baton Rouge, Louisiana 70803-4001
\end{instit}
\begin{abstract}
   We studied the finite temperature phase transition of SU(2)
gauge theory on four-dimensional Euclidean lattices by
Monte Carlo simulations, and measured the flux distributions of
a $q\bar q$ pair in both confined and unconfined phases. We
reviewed and generalized Michael sum rules to include
finite temperature effects. To compare our flux data with
predictions of Michael sum rules, we studied the behavior of
string tension with temperature. Our data agree well with the
generalized sum rules.

\end{abstract}

\narrowtext
\section{Introduction}
\label{sec: intro}

   The finite temperature phase transition of QCD~\cite{ps}
has been studied extensively in lattice gauge theory (LGT).
It has been shown that at low temperatures QCD is in the
confined phase, quarks are confined by a linear interquark
potential, $V(r)\sim \kappa r$, with $\kappa$ the string
tension. However, at sufficiently high temperatures the
system is in the unconfined phase.
Quarks are unconfined because the interquark potential is
a screened Coulomb potential~\cite{kt,eks,ro},
$V(r)\sim e^{-mr}/r $.

   It is believed that quark confinement is due to
the string formation in the confined phase, that is,
the color field between quarks forms a string-like
flux tube. In LGT one can probe the spacial distribution
of the energy density between a $q\bar q$ pair by using
a local operator (e.g., plaquette). Some strong numerical
evidence for string formation has been found~\cite{ph,wh,ha,so}.
Here we measured the flux data at various
temperatures. The color flux energy and action of the
$q\bar q$ pair are related to the interquark potential energy
$V(r)$ by the Michael sum rules~\cite{mis}.
In this paper our aim is to check these sum rules with
our finite temperature flux data.

   The Michael sum rule for energy is the LGT version of
the relation in classical electrostatics in which the work
to assemble a charge distribution is equal to the integral
over all space of the electrostatic energy density.
The Michael sum rule for energy and a similar relation
for action were tested on large lattices by Haymaker and
Wosiek~\cite{hw}. Except for an unresolved discrepency
arising in the determination of self energies, both sum
rules were satisfied by the flux data. In this paper we
rederive the sum rules for the case of finite temperature
and check them with our flux data.

   In the following we shall only consider pure SU(2) gauge
theory. To estimate the prediction of the Michael sum rules
at finite temperatures we need to know the lattice
scaling relation $a(\beta)$ and the behavior of the
string tension with temperature, $\kappa (T)$.
For this purpose we adopted the lattice asymptotic
scaling relation $a(\beta)$~\cite{cr} to the
non-perturbative region by allowing the scale constant to
vary slowly with $\beta$. Also the string tension was
measured at finite temperature, and it was found that the
string tension data agree very well with the fitting function,
\begin{equation}
 \kappa(T)=\kappa_0(1-\frac{T}{T_c})^\alpha
 \qquad \mbox {(for $T<T_c$)}.
\label{e1.1}
\end {equation}

   This paper is organized as follows. In Sec.~\ref{sec: thb}
we review some basic properties of the finite
temperature phase transition for SU(2) gauge theory.
The scaling relation $a(\beta)$ is studied, and the
flux measurement techniques in LGT is presented.
In Sec.~\ref{sec: mea} we describe our Monte Carlo
calculations, and present some numerical results for global
quantities, then we analyze the behavior of string tension
$\kappa$ with temperature. In Sec.~\ref{sec: mic}
we present a complete derivation for Michael sum rules
and generalize them to include finite size effects,
then we give a detailed comparison of these sum rules with
the flux data. Finally Sec.~\ref{sec: sum} presents the summary
and conclusions.

\bigskip
\section{Theoretical Background}
\label{sec: thb}

   In this section we present the fundamental ideas
of LGT in studying the finite temperature phase transition
of QCD, and emphasize aspects that are relevant for our
numerical investigations.

\subsection{ SU(2) Finite Temperature Phase Transition }
\label{subsec: ftp}

   In our study of SU(2) LGT we used the standard
Wilson action~\cite{cr}
\begin{equation}
 S(U)=\beta \sum_P(1-\frac{1}{2} {\rm Tr} U_P),
\label{e2.0a}
\end {equation}
  where $\beta=4/g^2$ with $g$ to be the lattice coupling
constant, and $U_P$ is the product of link variables around a
plaquette,
\begin{equation}
 U_P(n)=U_{\mu}(n)
U_{\nu}(n+\mu)U_{\mu}^{-1}(n+\nu)U_{\nu}^{-1}(n),
\label{e2.0b}
\end {equation}
  where the indices $\mu, \nu$ represent the orientation
of the plaquette, $n$ the position
of the plaquette on the lattice.

  On a four-dimensional Euclidean lattice of the
size, $N_t\times N_s^2\times N_z$, the temperature $T$ is
related to the temporal extent of the lattice~\cite{cb}.
If one chooses $N_t$ as the time direction, then one can
define the temperature as,
\begin{equation}
 T=\frac{1}{N_t a},
\label{e2.1}
\end {equation}
where $a$ is the lattice spacing, which is a function of
$\beta$.

   To study the finite temperature phase transition a convenient
order parameter is provided by the Polyakov loop closed in the
time direction~\cite{ct},
\begin{equation}
 P({\bf r})=\frac{1}{2} {\rm Tr}\prod_{\tau =1}^{N_t}U_t({\bf r},\tau),
\label{e2.2}
\end {equation}
   where $U_t({\bf r}, \tau)$ is the link variable oriented in
$N_t$ direction at the site $({\bf r}, \tau)$.
 It is well-known for pure SU($N$) gauge theory that the
expectation value of the Polyakov loop $\langle P\rangle$
plays the role of an order parameter~\cite{ct} in the finite
temperature phase transition, that is, in the infinite volume
limit one has

   in the confined phase, $T<T_c$, $\langle P\rangle =0$;

   in the unconfined phase, $T>T_c$, $\langle P\rangle \ne 0$;
\newline
   because of the spontaneous breaking of the global $Z_N$
symmetry~\cite{ms}. Here $T_c$ is the transition temperature.

 The expectation value $\langle P\rangle$ is associated with the
free energy $F_q$ of an isolated quark $q$ in the infinite volume
limit~\cite{ms},
\begin{equation}
 |\langle P\rangle|\sim e^{-L_t F_q},
\label{e2.3}
\end {equation}
  where $L_t=N_t a$, is the inverse temperature. Further, the
correlation function $\langle P(0)P^{\dagger}(r)\rangle$
is related to the potential energy $V(r)$ of a $q\bar q$ pair
in the infinite volume limit,
\begin{equation}
 \langle P(0)P^{\dagger}(r) \rangle \sim const. \exp[-L_t V(r)].
\label{e2.4}
\end {equation}
  In the confined phase, $T < T_c$, Eq.~(\ref{e2.4}) becomes,
for large $r$~\cite{kt},
\begin{eqnarray}
 \lim_{r\rightarrow \infty}\langle P(0)P^{\dagger}(r) \rangle
&\sim&  \exp[-L_t \kappa r], \nonumber \\
&=& \langle P \rangle ^2=0,
\label{e2.5}
\end {eqnarray}
because of $\langle P \rangle=0$ in the confined phase, where
$\kappa$ is the string tension. Eq.~(\ref{e2.5}) implies that
the $q\bar q$ potential in the confined phase is linear,
e.g., $V(r)\sim \kappa r$. On the other hand,
  in the unconfined phase, $T > T_c$, and in the infinite
volume limit it is found that for SU(2) gauge theory~\cite{kt},
\begin{eqnarray}
 \lim_{r\rightarrow \infty}\langle P(0)P^{\dagger}(r) \rangle
  &=& const.
[1+ L_t\frac{3}{\beta 4\pi r} \exp(-mr)] \nonumber \\
 & = &\langle P \rangle ^2\ne 0,
\label{e2.6}
\end {eqnarray}
for $\langle P \rangle \neq 0$ in the unconfined phase.
 Eq.~(\ref{e2.6}) implies that
the $q\bar q$ potential in the unconfined phase is a screened
Coulomb potential, e.g., $V(r)\sim \frac {e^{-mr}}{r}$ with
$m^{-1}$ the Debye screening length.
To the lowest order of perturbation theory one has~\cite{kt}
\begin{equation}
 m^2=\frac{2}{3}g^2 T^2.
\label{e2.7}
\end {equation}
Ref.~\cite{eks} gives
a detailed discussion about the $q\bar q$ potential
in the unconfined phase. We shall not discuss it any more
in the following.

  The transition temperature $T_c$ can be determined by Monte
Carlo calculations of LGT. For SU(2) one recent result
is~\cite{efw,em},
\begin{equation}
\beta_c = 2.2985\pm 0.0006.  \qquad \mbox{(for $N_t=4$)}
\label{e2.8}
\end {equation}
  This transition point $\beta_c$ was calculated on large
lattices (e.g., $4\times 26^3$), on which finite size effects
are small, the results can be considered to be
the transition point in the infinite volume limit.

   To obtain the transition temperature $T_c$ in physical units,
one must know the scaling relation between the lattice spacing
$a$ and $\beta$. If we assume the scaling relation $a(\beta)$ is
given by the asymptotic relation for SU(2)~\cite{cr},
\FL
\begin{equation}
 a(\beta)=
\Lambda_L^{-1}(\frac{6}{11}\pi^2\beta)^{\frac{51}{121}}
exp[-\frac{3}{11}\pi^2\beta]
=\Lambda_L^{-1}f(\beta),
\label{e2.9}
\end {equation}
where $\Lambda_L$ is the lattice scaling constant,
then one can calculate the transition temperature $T_c$ in
physical units from Eqs.~(\ref{e2.1}),~(\ref{e2.8})
and (\ref{e2.9}),
\begin{eqnarray}
 T_c\Lambda_L^{-1} &=& \frac{1}{N_t f(\beta_c)}, \nonumber \\
           &=& 42.11\pm 0.06.
\label{e2.10}
\end {eqnarray}
  This is consistent with the results of Refs.~\cite{bb,en}.
{}From Eqs.~(\ref{e2.9}) and (\ref{e2.10}) one can also see
that $\beta < \beta_c$ implies $T < T_c$,
and $\beta > \beta_c$ corresponds to $T > T_c$ for fixed $N_t$.
Now we proceed to study the lattice scaling constant
$\Lambda_L$.

\subsection{ The Scaling Relation $a(\beta)$ }
\label{subsec: ast}

   We follow Refs.~\cite{phm,mi} in choosing a scale such that
in the limit of zero temperature and infinite volumes the
string tension is
\begin{equation}
 \sqrt{\kappa_0} = 0.44 \quad \mbox{GeV},
\label{e4.1}
\end {equation}
which was determined from the Regge trajectory of
experimental data. Here for definiteness we use the
data from the real world to discuss SU(2) gauge
theory.

   From Eq.~(\ref{e4.1}) and the string tension data of
Monte Carlo calculations in Refs.~\cite{so,phm} one can extract
a relation between the lattice spacing $a$
and lattice coupling constant $\beta$. The result is
listed in Table~\ref{t4.1}, which is consistent with the
similar result of Ref.~\cite{hw}. In Table~\ref{t4.1} the values
of $\Lambda_L^{-1}$ were calculated
from the asymptotic scaling relation in Eq.~(\ref{e2.9}) by
substituting the corresponding values of $a$ and $\beta$
in this table. One can see
that the values of $\Lambda_L^{-1}$ changes slowly
with $\beta$ in the region, $2.22\le \beta \le 2.5$.
This implies that the perturbative asymptotic scaling relation
in Eq.~(\ref{e2.9}) is not exactly valid in this region.
To study the non-perturbative physics and obtain the
relation $a(\beta)$ for other $\beta$ values in this region, we
use the function in Eq.~(\ref{e2.9}) to fit the data of
Table~\ref{t4.1}, with $\Lambda_L^{-1}$ considered as a
function of $\beta$. For simplicity, we chose a quadratic
function to fit the $\Lambda_L^{-1}$ values. The result is
\begin{eqnarray}
 && a(\beta)= \Lambda_L^{-1}(\beta)f(\beta),  \nonumber \\
 \mbox{with} \qquad &&
\Lambda_L^{-1}(\beta)=d_1+d_2 \beta + d_3 \beta ^2 \quad \mbox{(fm)},
\label{e4.2}
\end {eqnarray}
where the coefficients $d_1=59.37\pm 0.86$, $d_2=-5.96\pm 0.39$
and $d_3=-3.28\pm 0.17$, and the function $f(\beta)$ is
given by Eq.~(\ref{e2.9}).

  Using the scaling relation of Eq.~(\ref{e4.2}) we can
transform the Monte Carlo data from lattice units to physical
units. Now the transition temperature $T_c$ in Eq.~(\ref{e2.10})
can be expressed explicitly in physical units, i.e.,
\begin{equation}
T_c=1.487\pm 0.140 \quad  \mbox{(1/fm)}.
\label{e4.3}
\end {equation}
Also, by using the scaling relation one can estimate the
transition point $\beta_c$ on lattices of any temporal size $N_t$,
if we assume the transition temperature $T_c$
in Eq.~(\ref{e4.3}) is independent of the lattice size $N_t$
because it is a physical observable.
For example, the estimated values of $\beta_c$ for $N_t=6$ and
8 are
\begin{eqnarray}
 \beta_c\simeq 2.42 &&\qquad \mbox {(for $N_t=6$)}, \nonumber \\
 \beta_c\simeq 2.50 &&\qquad \mbox {(for $N_t=8$)}.
\label{e4.4}
\end {eqnarray}
  In Sec.~\ref{sec: mic} we shall use the scaling relation of
Eq.~(\ref{e4.2}) to discuss the predictions of Michael sum rules.
In the following we proceed to describe the methods used in
measuring the $q\bar q$ flux distributions.

\subsection{$q\bar q$ Flux Distributions }
\label{subsec: mfd}

   To measure the flux distributions of a $q\bar q$ pair, we
calculated the quantity~\cite{wh,so},
\FL
\begin{equation}
 f_{\mu \nu}({\bf r}, {\bf x})=\frac {\beta}{a^4}
\biggl [\frac{\langle P(0)P^{\dagger}({\bf r})\Box_{\mu \nu}\rangle}
{\langle P(0)P^{\dagger}({\bf r})\rangle}
-\langle \Box_{\mu \nu}\rangle \biggr ],
\label{e3.4}
\end {equation}
   where the Polyakov loop $P({\bf r})$ is defined in
Eq.~(\ref{e2.2}), and $\Box_{\mu \nu}=\frac {1}{2} {\rm Tr}(U_P)$
is the plaquette variable with $U_P$ defined in Eq.~(\ref{e2.0b}).
The plaquette has 6 different orientations,
 $(\mu,\nu)$=$(2,3)$, $(1,3)$, $(1,2)$, $(1,4)$, $(2,4)$, $(3,4)$.

    To reduce the fluctuations of the quantity
$P(0)P^{\dagger}({\bf r})\Box$, in practical calculations we
measure the quantity~\cite{wh}
\FL
\begin{equation}
 f'_{\mu \nu}({\bf r}, {\bf x})=\frac {\beta}{a^4} \biggl
[\frac{\langle P(0)P^{\dagger}({\bf r})\Box_{\mu \nu}({\bf x})\rangle
- \langle P(0)P^{\dagger}({\bf r})\Box_{\mu \nu}({\bf x}_R)\rangle}
{\langle P(0)P^{\dagger}({\bf r})\rangle} \biggr ],
\label{e3.5}
\end{equation}
as the flux distribution instead of Eq.~(\ref{e3.4}),
where the reference point ${\bf x}_R$ was chosen to be
far from the $q\bar q$ sources. This
replacement does not change the measured average due to the
cluster decomposition theorem. The six components of
$f'_{\mu \nu}$ in Eq.~(\ref{e3.5}) correspond to
the components of the chromo-electric and chromo-magnetic fields
$({\bf \cal E}, {\bf \cal B})$ in Minkowski space, i.e.,
\begin{equation}
 f'_{\mu \nu}\rightarrow \frac {1}{2}
(-{\cal B}_1^2, -{\cal B}_2^2, -{\cal B}_3^2, {\cal E}_1^2,
{\cal E}_2^2, {\cal E}_3^2).
\label{e3.6}
\end{equation}
The total electric and magnetic energy densities are defined as
\begin{eqnarray}
 \rho_{\rm el} &=&\frac{1}{2}\langle {\cal E}^2\rangle
   =\frac{1}{2}[\langle{\cal E}_1^2\rangle+
\langle {\cal E}_2^2\rangle+\langle{\cal E}_3^2\rangle], \nonumber \\
 \rho_{\rm ma} &=&\frac{1}{2}\langle{\cal B}^2\rangle
  =\frac{1}{2}[\langle{\cal B}_1^2\rangle+
\langle{\cal B}_2^2\rangle+\langle{\cal B}_3^2\rangle].
\label{e3.7}
\end {eqnarray}

 To improve the statistical accuracy we used the multihit
techniques~\cite{wh}, that is, we make the replacement for Polyakov
loops in Eq.~(\ref{e3.5}),
\begin{equation}
 P({\bf r})\rightarrow  {\bar P}({\bf r})=
\frac {1}{2} {\rm Tr}\prod_{\tau =1}^{N_t}{\bar U}_t({\bf r},\tau),
\label{e3.8}
\end {equation}
where ${\bar U}_t({\bf r},\tau)$ is given by
\begin {eqnarray}
{\bar U}_t &=& \frac {\int U_t\exp \bigl
[\beta {\rm Tr} (U_tX_t^{\dagger}+h.c.)\bigr]dU_t}
{\int \exp \bigl[\beta {\rm Tr}(U_tX_t^{\dagger}+h.c.)\bigr]dU_t} ,
\nonumber \\
&=& X_t \frac {I_2(\beta\lambda)}
{\lambda I_1(\beta\lambda)},
\label{e3.9}
\end {eqnarray}
with $X_t^{\dagger}$ to be the neighborhood of $U_t$,
i.e., $U_t X_t^{\dagger}=\sum \Box_P$, the sum extends
over all plaquettes containing $U_t$. And
$\lambda=\sqrt {det(X_t)}$, $I_1$ and $I_2$ are the
modified Bessel functions. This technique is limited to
the situations where the neighboring links $X_t$ of
Polyakov loops $P$ do not overlap with the plaquette
$\Box_{\mu \nu}$ in Eq.~(\ref{e3.5}).

\bigskip

\section{ Numerical Results For Global Quantities }
\label{sec: mea}

  In this section we present the numerical results and
analyses for global quantities in our Monte Carlo calculations.
  We follow Ref.~\cite{bb} to choose our lattices of the
geometry $N_t\times N_s^2\times N_z$,
with $N_t\ll N_s\ll N_z$. If $N_t$ is chosen as the
time direction, we can simulate SU(2) gauge theory
at finite temperatures ($T=1/N_t a$) in large
volumes ($V=N_s^2 N_z a^3$).
  Typically, we choose $N_t=4$ and $6$, $N_s=7, 9, 11$, and
$N_z=65$ in most cases and $N_z=37$ in a few cases with $N_t=6$.
The lattice coupling constant $\beta$ is in the range
$2.25\le \beta \le 2.40$ for $N_t=4$, and
$2.30\le \beta \le 2.50$ for $N_t=6$.
We updated the lattice configurations by using the standard
Metropolis algorithm alternated with overrelaxation method.
We typically thermalized for 4000 sweeps, and made one
measurement every 10 sweeps. The total number of measurements
for each data is about 2000. The actual number may vary
by a small amount in each case. The calculations were
done on LSU's IBM 3090 mainframe.

\subsection{ Measurements of The Order Parameter $\langle |P|\rangle$ \\
    and $\langle P(0)P^{\dagger}(z)\rangle $ }
\label{subsec: oppp}

   As we discussed in Sec.~\ref{subsec: ftp} the
expectation $\langle P \rangle$ plays
the role of an order parameter in the infinite volume limit.
However, on a finite lattice this quantity is always
zero if the computation time is taken to be infinite,
because the system would flip between two ordered
states, and the values of the Polyakov loop $P$
would flip sign after some iterations. Therefore,
we take the expectation value of the modulus of the Polyakov
loop $\langle |P|\rangle$ as the ``order parameter" on finite
lattices~\cite{ct,ms}.

  The measured data of $\langle |P|\rangle$ are plotted in
Fig.~\ref{f3.1}, which shows a rapid increase
of $\langle |P|\rangle$ at $\beta\approx 2.30$ for $N_t=4$,
and another at $\beta\approx 2.42$ for $N_t=6$.
This implies that a phase transition occurs at
$\beta_c\approx 2.30$ for $N_t=4$, or $\beta_c\approx 2.42$
for $N_t=6$, in the infinite volume limit, which are
consistent with Eqs.~(\ref{e2.8}) and (\ref{e4.4}).
For cases of $N_t=4$ the data measured on the lattice
$4\times 9^2 \times 65$ agree very well with those from
$4\times 11^2 \times 65$, both approach the infinite volume limit.
However, for cases of $N_t=6$ the data from the lattice
$6\times 7^2 \times 65$ have some discrepencies with the
data from $6\times 11^2 \times 37$. This may be due to
finite-volume effects, as discussed in Ref.~\cite{efw},
because in these cases $N_s$ (or $N_z$) are not large
enough to approach the infinite volume limit.

   We also measured the correlations of two Polyakov loops,
$\langle P(0)P^{\dagger}(z)\rangle$, with the separation $z$
along $N_z$ direction. Some typical results are shown in
Figs.~\ref{f3.2} and~\ref{f3.3}, both were drawn as a logarithmic
plot. Fig.~\ref{f3.2} shows $\langle P(0)P^{\dagger}(r)\rangle$
versus $r$ in the confined phase, from this figure one can see
that the correlation $\langle P(0)P^{\dagger}(r)\rangle
\rightarrow 0$ as $r\rightarrow \infty$,
as predicted by Eq.~(\ref{e2.5}). Also, the linear behavior
of the quantity $\ln \langle P(0)P^{\dagger}(r)\rangle$ with $r$
indicates that the $q\bar q$ potential $V(r)$ is linear in the
confined phase. And the slope of the straight lines corresponds
to the string tension $\kappa$. This figure shows that the string
tension decreases with $\beta$ (or temperature). Fig.~\ref{f3.3}
is similar to Fig.~\ref{f3.2}, but is in the unconfined phase.
It shows that the correlation
$\langle P(0)P^{\dagger}(r)\rangle \not\rightarrow 0$
as $r\rightarrow \infty$, as predicted by Eq.~(\ref{e2.6}).

\subsection{ Measurements of String Tension $\kappa$ }
\label{subsec: mst}

      The string tension $\kappa$ can be extracted from
the correlation $\langle P(0)P^{\dagger}(r)\rangle$ according
to Eq.~(\ref{e2.5}). In our measurements we choose the
correlation along the longest extent $N_z$, i.e.,
${\bf r}=r {\bf e}_z$, and follow Ref.~\cite{bb} to extract the
string tension data from the relation,
\begin{equation}
 \langle P(0)P^{\dagger}(z)\rangle _c=const. (e^{-L_t \kappa z}+
e^{-L_t \kappa (L_z -z)}),
\label{e3.3}
\end {equation}
 where on the L.H.S. of this equation the connected
correlation $\langle P(0)P^{\dagger}(z)\rangle _c$ is defined as,
$\langle P(0)P^{\dagger}(z)\rangle -\langle P \rangle ^2$.
So the dependence on the
coordinates $t, x, y$ can be suppressed, and one can extract
the effective string tension. On the R.H.S. the new term
$e^{-L_t \kappa (L_z -z)}$ accounts for the finite size
effects along the correlation direction $N_z$,
with $L_z=N_z a$.

 The raw string tension data are given in Table~\ref{t3.2}.
Some data were measured in the transition
region, e.g., $\beta=2.30$ for ($N_t=4$).
In this case we assume Eq.~(\ref{e3.3}) be still valid.
Since our lattices have the geometry, $N_z\gg N_s\gg N_t$,
we find that finite size effects are small as $N_s/N_t \ge 2$
in the limit of $N_z\rightarrow \infty$. The detailed
analysis about the finite size effects is presented
elsewhere~\cite{ph}. This conclusion also agree with the
results of Refs.~\cite{bb,bbs}. Therefore, we expect that
the data measured on lattices $4\times 9^2\times 65$ and
$4\times 11^2\times 65$ can be viewed as the string tension
in the large volume limit. However, the data from
the lattice $6\times 7^2\times 65$ might have some large
finite size effects because $N_s=7$ is not large enough
for $N_t=6$. For data measured on the lattice
$6\times 11^2\times 37$ the transverse size $N_s=11$
is large, but some large finite size effects
may be caused by the relatively small longitudinal size $N_z$.

\subsection{ Behaviour of $\kappa$ With Temperature }
\label{subsec: stc}

  In Sec.~\ref{sec: mic} when we discuss Michael sum rules
and analyze the finite temperature effects on the flux data,
we need to know the relation between string tension $\kappa$
and temperature. In this part we proceed to analyze the
string tension data.

  Using the scaling relation in Eq.~(\ref{e4.2}) we can
transform the string tension data in
Table~\ref{t3.2} from lattice units to physical units.
The results are shown in Table~\ref{t4.2}.
{}From this table one can see that the string tension $\kappa$
decreases with temperature $T$ (or $\beta$) in the
confined phase. If we assume that $\kappa$ is a continous
function of $T$ (or $\beta$), one has
\begin{eqnarray}
 \frac{\partial \kappa}{\partial T} < 0 &&\qquad
\mbox{(for $T<T_c$)}, \nonumber \\
\mbox{or} \quad
\frac{\partial \kappa}{\partial \beta} < 0 &&\qquad
\mbox{(for $\beta<\beta_c$)}.
\label{e4.5}
\end {eqnarray}
   This behavior is also confirmed in other recent
papers~\cite{de,en}.

  Since the string tension $\kappa$ decreases with
temperature $T$ in the confined phase and is expected
to vanish at the transition point $T_c$, for simplicity,
we then use the simple function with power behavior
in Eq.~(\ref{e1.1}) to fit the string tension near $T_c$,
that is,
\begin{equation}
 \kappa(T)=\kappa_0(1-\frac{T}{T_c})^\alpha \qquad
\mbox {(for $T<T_c$)}.
\label{e4.6}
\end {equation}
  This is a simplified form of the fitting function used in
Ref.~\cite{en}. Here the constant $\kappa_0$ represents
the string tension $\kappa$ in the limit of zero temperature
and infinite volumes, which can be given by Eq.~(\ref{e4.1}),
i.e., $\kappa_0 \approx 0.981$~{GeV/fm}. The transition temperature
$T_c$ is given by Eq.~(\ref{e4.3}) in physical units.

   As we discussed above, the string tension data measured
on lattices $6\times 7^2\times 65$ and
$6\times 11^2\times 37$ may have large finite size effects.
By viewing the data in Table~\ref{t4.2} we find that the
string tension data from the lattice $6\times 7^2\times 65$
are consistent with the data from lattices $4\times 9^2\times 65$
and $4\times 11^2\times 65$, on which finite size effects can
be neglected, as we discussed before. However, data from the
lattice $6\times 11^2\times 37$ have large discrepancies with
other data. This suggests that finite size effects
on string tension are sensitive to the longitudinal
size $N_z$, but not as much to the transverse size $N_s$.

  We then use the string tension data from lattices of
small finite size effects (i.e., $4\times 11^2\times 65$
and $6\times 7^2\times 65$), to fit the function
in Eq.~(\ref{e4.6}).
The only fitting parameter is the exponential index $\alpha$,
the best fit result is,
\begin{equation}
 \alpha=0.35\pm 0.04.
\label{e4.7}
\end {equation}
  Fig.~\ref{f4.1} shows this fitting function and the measured
string tension data of small finite size effects.
In this figure one can see that a few data point very close to
the transition point $T_c$ have large discrepancies from
the fitting function, because fluctuations are large in the
transition region. We then excluded these data at,
$T \approx 1.48$ (1/fm), from our
fitting process. Also all other data above $T_c$ were excluded
because they would correspond to complex values for
the fitting function. Below the transition region our
data agree very well with the conjectured behavior in
Eq.~(\ref{e4.6}).

  To get the relation between $\kappa$ and $\beta$,
we also fit the string tension data with the following function
similar to Eq.~(\ref{e4.6}),
\begin{equation}
 \kappa(\beta)=\kappa_0(1-\frac{\beta}{\beta_c})^\delta
\qquad \mbox {(for $\beta <\beta_c$)}.
\label{e4.8}
\end {equation}
The transition point $\beta_c=2.2985\pm 0.0006$ for $N_t=4$
can be chosen from Eq.~(\ref{e2.8}), and $\beta_c=2.42\pm 0.01$
for $N_t=6$ from Eq.~(\ref{e4.4}). We find that the best fit
results for the exponential index $\delta$ are
\begin{eqnarray}
 \delta =0.22\pm 0.03 &&\qquad \mbox{(for $N_t=4$)}, \nonumber \\
 \delta =0.14\pm 0.04 &&\qquad \mbox{(for $N_t=6$)}.
\label{e4.9}
\end {eqnarray}
  In Fig.~\ref{f4.2} we plot the result for the case of
$N_t=4$. Also, in this figure the data above the transition
point, $\beta_c = 2.2985$, were excluded from our fitting
process. Again our data agree well with the fitting
function. In Sec.~\ref{subsec: afd} we shall use the
relation $\kappa(\beta)$ to check the Michael sum rules.

\bigskip
\section{Michael Sum Rules And Flux Distributions }
\label{sec: mic}

   To study the detailed properties of the finite temperature
phase transition, we measured the $q\bar q$ flux distributions
at various temperatures using the technique discussed in
Sec.~{subsec: mfd}. The flux energies (action)
are related to the potential energy $V(r)$ of the
$q\bar q$ pair by Michael sum rules, which were first
derived by C. Michael~\cite{mis}. In the derivation
of these sum rules some scaling relations on asymmetric
lattices are used, which were studied extensively by
A. Hasenfratz and P. Hasenfratz~\cite{hh}, and F. Karsch~\cite{ka}.
The original work of C. Michael is somewhat terse and
drawing on other references. In order to
generalize it we feel it is helpful to present a thorough
derivation, including the other references. In this section
 we shall give a complete derivation of
Michael sum rules, and present our supplements and
generalizations. One result is delegated to the Appendix.
Then we compare our flux data with the
predictions of these sum rules.

\subsection{ Review and Generalizations of Michael Sum Rules }
\label{subsec: mfs}

    For a static $q\bar q$ system at the spatial
separation $r$ the Michael sum rules states that,
in the limit of zero temperature, infinite volume and
infinitesimal lattice spacing ($a\rightarrow 0$),
the electric and magnetic flux energies of the system
in Minkowski space are~\cite{mis},
\FL
\begin{equation}
 E_{\rm el}(r)=\sum_s\frac{1}{2} a^3 \langle {\cal E}^2 \rangle
=-\frac {1}{2}V(r)[\frac{\partial \ln a}{\partial \ln \beta}-1]
-\frac {1}{2a}[\beta \frac{\partial f}{\partial \beta}-f],
\label{e5.1}
\end {equation}
\begin{equation}
 E_{\rm ma}(r)=\sum_s\frac{1}{2} a^3 \langle {\cal B}^2 \rangle
=\frac {1}{2}V(r)[\frac{\partial \ln a}{\partial \ln \beta}+1]
+\frac {1}{2a}[\beta \frac{\partial f}{\partial \beta}+f].
\label{e5.2}
\end {equation}
where the sums extends over the whole space around the
$q\bar q$ pair, $V(r)$ is the potential energy and $f/a$ is the self
energy of the $q\bar q$ sources. $E_{\rm el}(r)$ and $E_{\rm ma}(r)$
denote the electric and magnetic parts of the flux energy,
$\langle {\cal E}^2 \rangle$ and $\langle {\cal B}^2 \rangle$
are given by Eq.~(\ref{e3.7}).

\subsubsection{The Action Sum Rule}
\label{ssubsec: asr}

     To derive Eqs.~(\ref{e5.1}) and (\ref{e5.2}),
it is convenient to work in the transfer matrix formalism of LGT.
We choose the temporal gauge~\cite{ko},
that is, all temporal links are trivial,
$U_t(n)=e^{igaA_{\mu}(n)}=1$.
One remark about the temporal gauge of a finite lattice is that
along a time axis one can not choose all temporal links trivial,
$U_t=1$, because of the restriction that a Polyakov loop in the
time direction is gauge invariant, which has values other than
the trivial one, $P(\vec r)=1$. Therefore, on a finite
lattice one can choose all temporal links to be trivial,
except one link on each time axis, which can be chosen
far from the operators under considerations.

    Let's consider a Wilson loop $W$ of the temporal size $na$ and
the spatial size $r$, as shown in Fig.~\ref{f5.1}. The time
directed pathes represent the $q\bar q$ sources, the space directed
pathes $P_r(0)$ and $P_r(n)$ create and annihilate the static
$q\bar q$ pair from and to the vacuum. In the temporal gauge the
expectation of Wilson loop becomes,
\begin{equation}
 \langle W \rangle = \langle P_r(0) P_r(n)\rangle,
\label{e5.3}
\end {equation}
where the expectation is evaluated in the partition function form,
\begin{equation}
 \langle W \rangle =
\frac {\int d[U] e^{-\beta S'} W }{\int d[U] e^{-\beta S'}},
\label{e5.26}
\end {equation}
with $\beta S'=S$, the action given by Eq.~(\ref{e2.0a}),
i.e., $S'=\sum (1-\Box)$, where $\Box$ is defined in Eq.~(\ref{e3.4}),
the sum is over all plaquettes on the lattice.

  Using the transfer matrix approach~\cite{ko,ha} and assuming a
discrete spectrum for the lowest eigenstates of the transfer
matrix, one can evaluate Eq.~(\ref{e5.3}) as
\begin{eqnarray}
 \langle W \rangle &=& \frac {1}{Z} {\rm Tr}(P_r(0){\cal T}^n_{q\bar q}
 P_r(n){\cal T}^{N_t-n}) \nonumber \\
  &=& \frac {1}{\sum_\alpha \lambda_{\alpha}^{N_t}}
\sum_{\mu, \nu} \langle \mu | P_r|\nu, r \rangle \lambda^n_\nu (r)
\langle\nu, r| P_r|\mu\rangle  \lambda^{N_t-n}_\mu \nonumber \\
  &=& \frac {1}{\sum_\alpha \lambda_{\alpha}^{N_t}}
\sum_{\mu, \nu} d_{\mu \nu} d_{\nu \mu} \lambda^n_\nu (r)
\lambda^{N_t-n}_\mu,
\label{e5.4}
\end {eqnarray}
where ${\cal T}_{q\bar q}$ is the transfer matrix projected into
the $q\bar q$ sector of the Hilbert space~\cite{ha}, ${\cal T}$ is
the transfer matrix for the vacuum. $|\nu, r\rangle$ and
$|\mu\rangle$ are the eigenstates of ${\cal T}_{q\bar q}$ and
${\cal T}$ respectively, and $\lambda_\nu (r)$, $\lambda_\mu$ are
the corresponding eigenvalues, with $\lambda_\nu (r)=e^{-aE_{\nu}(r)}$
and $\lambda_\mu=e^{-aE_{\mu}}$. The coefficient
$d_{\mu \nu}=\langle \mu | P_r|\nu, r\rangle$. In the limit of
$N_t\rightarrow \infty$, the partition function $Z=\lambda_0^{N_t}$,
and in Eq.~(\ref{e5.4}) the dominant contributions correspond
to $\mu =0$, that is,
\begin{equation}
 \langle W \rangle =\sum_{\nu} d_{0 \nu}^2
\biggl(\frac {\lambda_\nu (r)}{\lambda_0} \biggr)^n
\qquad \mbox{(as $N_t \rightarrow \infty$ )}.
\label{e5.5}
\end {equation}
   If the temporal size $n$ of the Wilson loop $W$ is very large
($n\rightarrow \infty$), the dominant term in Eq.~(\ref{e5.5})
is given by $\nu=0$,
\begin{equation}
 \langle W \rangle =
 d_{0 0}^2 \biggl(\frac {\lambda_0 (r)}{\lambda_0} \biggr)^n
 =d_{0 0}^2 e^{-na(E_0(r) -E_0)},
\qquad \mbox{ (as $N_t$, $n \rightarrow \infty$ )}
\label{e5.6}
\end {equation}
where the energy of the vacuum $E_0$ is usually chosen to be zero,
in the following we will take this choice which implies
$\lambda_0=1$.

   Now we consider the $\beta$-derivative of Eq.~(\ref{e5.6}),
that is,
\begin{equation}
\frac{\partial \langle W \rangle }{\partial \beta}=
 -\langle WS' \rangle +\langle W \rangle \langle S' \rangle
 =\frac{\partial}{\partial \beta}
  [d_{0 0}^2 e^{-naE_0(r)}],
\label{e5.7}
\end {equation}
where we have taken $E_0=0$, and $S'$ is defined after
Eq.~(\ref{e5.26}).

  Consider a plaquette $\Box (m)$ outside the Wilson loop $W$
in the time direction (i.e., $0< n < m$), e.g., the plaquette
$P_1$ in Fig.~\ref{f5.1}, where we draw a plaquette with a
time extension. In the limit of infinite large $n$ and infinitesimal
lattice spacing, $a\rightarrow 0$, one can neglect the time
extension of $P_1$. The contribution of $P_1$
to the L.H.S. of Eq.~(\ref{e5.7}) is ${\langle W \Box (m)\rangle}
-{\langle W \rangle} {\langle \Box (m)\rangle}$.
As $\Box (m)$ is far from $W$ ( $m-n \gg 1$ ), one has
${\langle W \Box (m)\rangle} \approx {\langle W \rangle}
{\langle \Box (m)\rangle}$, then the contribution vanishes,
${\langle W \Box (m)\rangle} -{\langle W \rangle}
{\langle \Box (m)\rangle} \rightarrow 0$. However,
when $\Box (m)$ is close to $W$ ( $m\approx n$ ), one expects
that this contribution does not vanish. Now let us
show this explicitly,
\begin{eqnarray}
 && \langle W \Box (m)\rangle
    - \langle W \rangle \langle \Box (m)\rangle
\qquad \mbox{( for $0< n< m $ )} \nonumber \\
 && =\frac {1}{Z} {\rm Tr} (P_r(0){\cal T}^n_{q\bar q} P_r(n)
{\cal T}^{m-n} \Box (m){\cal T}^{N_t-m})
- \langle W \rangle \langle \Box (m)\rangle \nonumber \\
 && {\stackrel {N_t\rightarrow \infty}{\longrightarrow}}
\sum_{\mu, \nu} d_{0 \mu }
\biggl(\frac {\lambda_\mu (r)}{\lambda_0} \biggr)^n d_{\mu \nu}
\biggl(\frac {\lambda_\nu}{\lambda_0} \biggr)^{m-n}
\langle \nu |\Box (m)|0 \rangle \nonumber \\
 && \qquad \qquad -\sum_{\mu} d_{0 \mu}^2
\biggl(\frac {\lambda_\mu (r)}{\lambda_0} \biggr)^n
\langle 0|\Box (m)|0\rangle,
\label{e5.8}
\end {eqnarray}
where in the last step we have used Eq.~(\ref{e5.5})

  From Eq.~(\ref{e5.8}) one can see that, as $m-n\gg 1$,
the dominant term of $\langle W \Box (m)\rangle$
is given by $\nu=0$, which would be cancelled by
the product $\langle W \rangle \langle\Box (m)\rangle$.
So one has $\langle W \Box (m)\rangle -
\langle W \rangle \langle\Box (m)\rangle \approx 0$,
as $m-n\gg 1$.

   However, for $m\approx n$ the term of $\nu=0$ of the
quantity, $\langle W \Box (m)\rangle$, in Eq.~(\ref{e5.8})
is cancelled by the product
$\langle W \rangle \langle\Box (m)\rangle$. The major
contribution comes from the term of $\nu=1$. In the limit
of $n\rightarrow \infty$ Eq.~(\ref{e5.8}) becomes,
\begin{eqnarray}
&& \langle W \Box (m)\rangle
    - \langle W \rangle \langle \Box (m)\rangle
\qquad \mbox{( for $0< n< m $ )} \nonumber \\
 && \approx \sum_{\mu} d_{0 \mu }
\biggl(\frac {\lambda_\mu (r)}{\lambda_0} \biggr)^n
 d_{\mu 1} \biggl(\frac {\lambda_1}{\lambda_0} \biggr)^{m-n}
\langle 1|\Box (m)|0 \rangle \nonumber \\
 && \stackrel{n\rightarrow \infty}{\longrightarrow}
    e^{-naE_0(r)} d_{0 0 } d_{0 1} {\lambda_1}^{m-n}
\langle 1|\Box (m)|0 \rangle .
\label{e5.9}
\end {eqnarray}
where in the last step we have used the fact that $\lambda_0=1$,
and the dominant term is $\mu=0$ as $n\rightarrow \infty$.
Eq.~(\ref{e5.9}) implies that when the plaquette $\Box (m)$
is close to the Wilson loop in the time direction,
it gives a contribution of the order $e^{-naE_0(r)}$,
because the coefficients $d_{0 0 }$, $d_{0 1}$ and
$\lambda_1^{m-n}$ are of order of unity in this case.

  Finally for the plaquette $\Box (m)$
inside the Wilson loop $W$ in the time direction
(i.e., $0< m < n$), e.g., the plaquette $P_2$ in
Fig.~\ref{f5.1}, its contribution to the L.H.S. of
Eq.~(\ref{e5.7}) is,
\begin{eqnarray}
&& \langle W \Box (m)\rangle
    - \langle W \rangle \langle \Box (m)\rangle
 \qquad \mbox{( $0< m< n $ )} \nonumber \\
&& =\frac {1}{Z} {\rm Tr} (P_r(0){\cal T}^m_{q\bar q} \Box (m)
{\cal T}_{q\bar q}^{n-m} P_r(n){\cal T}^{N_t-n})
-\langle W \rangle \langle \Box (m)\rangle \nonumber \\
&& \stackrel{N_t\rightarrow \infty}{\longrightarrow}
\sum_{\mu, \nu} d_{0 \mu } d_{\nu 0}
\biggl(\frac {\lambda_\mu (r)}{\lambda_0} \biggr)^{m}
\biggl(\frac {\lambda_\nu (r)}{\lambda_0} \biggr)^{n-m}
\langle \mu,r|\Box (m)|\nu,r\rangle  \nonumber \\
&& \qquad \qquad -\langle W \rangle \langle \Box (m)\rangle .
\label{e5.10}
\end {eqnarray}
    In the limit of $n\rightarrow \infty$, and for cases
that both $m$ and $n-m$ are large, the dominant term of
Eq.~(\ref{e5.10}) is given by $\mu=\nu=0$, that is,
\begin{eqnarray}
 && \langle W \Box (m)\rangle
    - \langle W \rangle \langle \Box (m)\rangle
  \qquad \mbox{( $0< m< n $ )} \nonumber \\
 && \approx d_{0 0}^2
\biggl(\frac {\lambda_0 (r)}{\lambda_0} \biggr)^n
[\langle 0,r|\Box (m)|0,r \rangle-\langle 0|\Box (m)|0\rangle ]
    \nonumber \\
 && = d_{0 0}^2
\biggl(\frac {\lambda_0 (r)}{\lambda_0} \biggr)^{n}
(\Box )_{r-0},
\label{e5.11}
\end {eqnarray}
where we denote $(\Box )_{r-0}=[\langle 0,r|\Box (m)|0,r\rangle
-\langle 0|\Box (m)|0\rangle ]$, and Eq.~(\ref{e5.6}) is used.

  From the above discussion we can conclude that for plaquettes
outside the Wilson loop $W$ in the time direction, their
contributions to the L.H.S. of Eq.~(\ref{e5.7}) can be neglected
as they are far enough from $W$. For the Wilson loop of large
temporal size, i.e., $n\rightarrow \infty$, the major contribution
to the L.H.S. of Eq.~(\ref{e5.7}) comes from plaquettes inside
$W$. The contribution of one plaquette inside $W$ is given by
Eq.~(\ref{e5.11}). Summing over contributions from all such
plaquettes gives the dominant term of the L.H.S. in Eq.~(\ref{e5.7}),
that is,
\begin{eqnarray}
 \frac{\partial \langle W \rangle}{\partial \beta}
  &=& -\langle W S'\rangle + \langle W \rangle \langle S'\rangle
\nonumber \\
  &\approx & n d_{0 0}^2
\biggl(\frac {\lambda_0 (r)}{\lambda_0} \biggr)^n
(\sum_s\Box )_{r-0},
\label{e5.12}
\end {eqnarray}
where the sum $\sum_s$ is over all plaquettes in the spatial
volume at one fixed time value $m$ ( $0< m< n $ ). The
factor of $n$ comes from summing equal contributions
from plaquettes on each time slice.
  The dominant term of the R.H.S. in Eq.~(\ref{e5.7}) is,
as $n\rightarrow \infty$,
\begin{equation}
 \frac{\partial}{\partial \beta}
[d_{0 0}^2 e^{-naE_0(r)}]= -n d_{0 0}^2 e^{-naE_0(r)}
\frac{\partial [aE_0(r)]}{\partial \beta}.
\label{e5.13}
\end {equation}

   Collecting Eqs.~(\ref{e5.7}), (\ref{e5.12}) and (\ref{e5.13})
yields,
\begin{equation}
 -\frac{\partial aE_0(r)}{\partial \beta}=
 (\sum_s \Box )_{r-0},
\label{e5.14}
\end {equation}
where the R.H.S. can be measured by calculating the quantity,
$\langle W\Box \rangle /\langle W \rangle$ in LGT.
  In the continuum limit ($a\rightarrow 0$) the R.H.S. of
Eq.~(\ref{e5.14}) becomes,
\begin{equation}
 (\sum \Box )_{r-0}=
\frac{1}{\beta}\sum_s \frac{1}{2} a^4 [\langle {\cal E}^2
\rangle_{r-0}-\langle {\cal B}^2\rangle_{r-0}]
=\frac {a}{\beta}A,
\label{e5.15}
\end {equation}
where ${\bf \cal E}$ and ${\bf \cal B}$ are the color electric
and color magnetic fields in Minkowski space, $A$ is the
integration of the action density over the spatial volume.

   On the L.H.S. of Eq.~(\ref{e5.14}) if we write the
$q\bar q$ colour field energy $E_0(r)$ in the form~\cite{mis},
\begin{equation}
 E_0(r)= V(r) + f(\beta)/a,
\label{e5.16}
\end {equation}
where $V(r)$ is the potential energy, $f/a$ the self-energy.
Then one can obtain the following relation from
Eq.~(\ref{e5.14}) in the continuum limit,
\begin{equation}
 A =-\frac{\beta}{a}\frac{\partial aE_0(r)}{\partial \beta}
=-\frac{\beta}{a}\frac{\partial aV(r)}{\partial \beta}
-\frac{\beta}{a}\frac{\partial f}{\partial \beta}.
\label{e5.17}
\end {equation}
This is just the Michael's action sum rule,
which can be obtained by subtracting Eq.~(\ref{e5.2})
from Eq.~(\ref{e5.1}) because of $A=E_{\rm el}-E_{\rm ma}$,
if we assume that $V(r)$ scales and hence is independent
of $\beta$.

\subsubsection{Energy Sum Rules and Generalizations }
\label{ssubsec: esrg}

  To derive the energy sum rules in Eqs.~(\ref{e5.1}) and
(\ref{e5.2}) we need to study the color electric and color
magnetic fields separately. Let us consider an asymmetric
lattice with the time-spacing $a_t$, the spatial-spacing
$a_s$ $(=a)$ and the asymmetry $\xi =a_s/a_t $~\cite{mis}.
The action for SU($N$) LGT becomes,
\begin{equation}
 S_A = -\beta_t \sum \Box_t - \beta_s \sum \Box_s + const. ,
\label{e5.18}
\end {equation}
where $\Box_t$ is a plaquette with a time extent and $\Box_s$
is a space-like plaquette. In the continuum limit
($a_s$, $a_t\rightarrow 0$) the action must become the
classical action,
\begin{equation}
  S_A \longrightarrow \frac {1}{4}\int d^4x (F^c_{\mu \nu})^2,
\label{e5.19}
\end {equation}
 In this limit the time-like plaquette becomes
\begin{eqnarray}
 \beta_t \Box_t &&=\beta_t \Box_{j4}
=\beta_t \frac {1}{N}tr(e^{ia_t a_s g_t F_{j4}}) \nonumber \\
 &&\approx \beta_t (1-\frac {1}{4N}
 a_t^2 a_s^2 g_t^2 (F^c_{j4})^2 ),
\label{e5.20}
\end {eqnarray}
where $N$ is for SU($N$). $j=1,2,3$ and $c$ is the colour index.
  To get the correct continuum action, Eq.~(\ref{e5.19}),
we require that in Eq.~(\ref{e5.20}),
\[ \frac {\beta_t}{2N}a_t^2 a_s^2 g_t^2 =a_s^3 a_t, \]
from which we obtain the relation between $\beta_t$ and $g_t$.
\begin{equation}
  \beta_t=\frac {2N}{g_t^2} \frac{a_s}{a_t}
 =\frac {2N}{g_t^2} \xi .
\label{e5.21}
\end {equation}
  Similarly one can obtain the relation between $\beta_s$ and $g_s$,
\begin{equation}
  \beta_s=\frac {2N}{g_s^2} \frac{a_t}{a_s}
 =\frac {2N}{g_s^2} \xi ^{-1}.
\label{e5.22}
\end {equation}

 In the weak coupling limit, $\beta_s$, $\beta_t$ can be expanded in
terms of the coupling $\beta$ of the corresponding symmetric
lattice~\cite{mis,hh},
\begin{mathletters}
\begin{eqnarray}
 \beta_s \xi &=& \frac {2N}{g_s^2}
              =\beta (a) +2N c_s(\xi) +O(\beta^{-1}),
\label{e5.23a} \\
 \beta_t \xi^{-1} &=& \frac {2N}{g_t^2}
                   =\beta (a) + 2N c_t(\xi) + O(\beta^{-1});
\label{e5.23b}
\end{eqnarray}
\end{mathletters}
where the coefficients $c_s(\xi)$ and $c_t(\xi)$ satisfy the
conditions~\cite{mis,ka},
\begin{eqnarray}
 c_s(\xi)|_{\xi=1} &=& c_t(\xi)|_{\xi=1}=0, \nonumber \\
 \frac{\partial }{\partial \xi}
[c_s(\xi)+c_t(\xi)]_{\xi=1} &=& (c'_s + c'_t)_{\xi=1}
 =b_0= 11N/48{\pi^2}.
\label{e5.24}
\end {eqnarray}
where $c'_s$ ($c'_t$) denotes the $\xi$-derivative of $c_s$ ($c_t$).

 With these definitions we can now study the color electric and
magnetic fields of a $q\bar q$ sources. Again we consider a
Wilson loop $W$ of the size $n a_t$ in the time direction
and $r$ in the space direction on the asymmetric lattice.
We can obtain the following result in this case, if we
repeat the steps that lead to Eq.~(\ref{e5.6}),
\begin{equation}
 \langle W \rangle= d_{0 0}^2 e^{-n{a_t}E_0(r) },
\qquad \mbox{(as $N_t$, $n \rightarrow \infty$ )}
\label{e5.25}
\end {equation}
where $E_0 (r)$ is the ground-state energy of the $q\bar q$ pair,
and $\langle W \rangle$ is defined by the partition function formalism
in Eq.~(\ref{e5.26}), with the action $\beta S'$ replaced by
$S_A$ in Eq.~(\ref{e5.18}).

 By taking the derivatives of Eq.~(\ref{e5.25}) with respect to
$\beta_s$ and $\beta_t$ respectively, one can get the following
relations similar to Eq.~(\ref{e5.14}),
\begin{mathletters}
\begin{eqnarray}
 -\frac{\partial a_t E_0(r)}{\partial \beta_t}
 &=& (\sum_s \Box_t )_{r-0},  \label{e5.27a} \\
 -\frac{\partial a_t E_0(r)}{\partial \beta_s}
 &=& (\sum_s \Box_s )_{r-0}.
\label{e5.27b}
\end{eqnarray}
\end{mathletters}
where the sum $\sum_s$ is over the whole spacial volume as
before. The R.H.S. of Eqs.~(\ref{e5.27a}) and~(\ref{e5.27b})
become the total color electric and color magnetic
energies of the $q\bar q$ pair respectively in the continuum
limit ($a_s, a_t\rightarrow 0)$. On the symmetric lattice
($\xi =1$) one has,
\begin{mathletters}
\begin{eqnarray}
 \lim_{a\rightarrow 0} (\sum_s \Box_t)_{r-0} &=&
\frac{1}{\beta}\sum_s \frac {1}{2} a^4
 \langle {\cal E}^2\rangle_{r-0},
\label{e5.28a} \\
 \lim_{a\rightarrow 0} (\sum_s \Box_s)_{r-0} &=&
-\frac{1}{\beta}\sum_s \frac{1}{2} a^4
 \langle {\cal B}^2\rangle_{r-0};
\label{e5.28b}
\end{eqnarray}
\end{mathletters}
where ${\bf \cal E}$ and ${\bf \cal B}$ are the color electric
and magnetic fields in Minkowski space.

   To evaluate the L.H.S. of Eqs.~(\ref{e5.27a}) and (\ref{e5.27b})
we need to resort following relations which relate the quantities on
the asymmetric lattice with the quantities on the equivalent
symmetric lattice,
\begin{mathletters}
\begin{eqnarray}
 \biggl(\frac {\partial F(\beta_s, \beta_t)}
 {\partial \beta_t} \biggr)_{\xi=1} &=&\frac{1}{2} \biggl(
 \frac {\partial F}{\partial \beta}+\frac{1}{\beta}\frac
{\partial F}{\partial \xi} \biggr)_{\xi=1},
\label{e5.29a} \\
 \biggl(\frac {\partial F(\beta_s, \beta_t)}
 {\partial \beta_s} \biggr)_{\xi=1} &=&\frac{1}{2} \biggl(
 \frac {\partial F}{\partial \beta}-\frac{1}{\beta}\frac
{\partial F}{\partial \xi} \biggr)_{\xi=1} ;
\label{e5.29b}
\end{eqnarray}
\end{mathletters}
where $F(\beta_s, \beta_t)$ is a function of $\beta_s$, $\beta_t$.
The proof is given in the Appendix.

    Since we can write the energy $E_0(r)$ of Eqs.~(\ref{e5.29a})
and (\ref{e5.29b}) in the form, $ E_0(r)= V(r) + f(\beta)/a$,
from Eq.~(\ref{e5.16}), then the L.H.S. of Eq.~(\ref{e5.27a})
becomes,
\begin{equation}
 \biggl(\frac{\partial a_t E_0(r)}
{\partial \beta_t}\biggr)_{\xi=1}=\biggl(\frac{\partial a_t V(r)}
{\partial \beta_t}\biggr)_{\xi=1}+\biggl(\frac{\partial (f/\xi)}
{\partial \beta_t}\biggr)_{\xi=1} .
\label{e5.30}
\end {equation}
  Applying Eq.~(\ref{e5.29a}) to above equation, and using the
fact that the potential $V(r)$ is independent of $\beta$ in the
limit of infinite lattice sizes and infinitesimal lattice spacing
($a\rightarrow 0$), one can obtain,
\begin{eqnarray}
 \biggl(\frac{\partial a_t V(r)}
{\partial \beta_t}\biggr)_{\xi=1} &=& \frac {1}{2}\biggl(
\frac{\partial a_t V(r)}{\partial \beta} + \frac{1}{\beta}
\frac{\partial a_t V(r)}{\partial \xi}\biggr)_{\xi=1} \nonumber \\
 &=& \frac {1}{2} V(r)\biggl(
\frac{\partial a}{\partial \beta} - \frac{a}{\beta} \biggr),
\label{e5.31}
\end{eqnarray}
where we have used $\bigl(\frac{\partial a_t}
{\partial \beta}\bigr)_{\xi=1}=\frac{\partial a}
{\partial \beta}$ and $\bigl(\frac{\partial a_t}{\partial \xi}
\bigr)_{\xi=1}=-a $. Also the second term on the L.H.S. of
Eq.~(\ref{e5.30}), $\bigl(\frac{\partial (f/\xi)}
{\partial \beta_t}\bigr)_{\xi=1}$, can be evaluated as,
\begin{equation}
 \biggl(\frac{\partial (f/\xi)}
{\partial \beta_t}\biggr)_{\xi=1}= \frac {1}{2}\biggl(
\frac{\partial f(\beta)}{\partial \beta} - \frac{f(\beta)}
{\beta}\biggr).
\label{e5.32}
\end {equation}
where the self-energy $f(\beta)/a$ depends on $\beta$ and $a$.
  Substituting Eqs.~(\ref{e5.31}) and~(\ref{e5.32}) into
Eq.~(\ref{e5.30}) yields
\FL
\begin{equation}
 \biggl(\frac{\partial a_t E_0(r)}
{\partial \beta_t}\biggr)_{\xi=1}=\frac {1}{2} V(r)\biggl(
\frac{\partial a}{\partial \beta} - \frac{a}{\beta} \biggr)
+\frac {1}{2}\biggl(\frac{\partial f(\beta)}{\partial \beta}
-\frac{f(\beta)}{\beta}\biggr).
\label{e5.33}
\end {equation}

  Similarly, the L.H.S. of Eq.~(\ref{e5.27b}) can be evaluated
by applying Eq.~(\ref{e5.29b}), the result is,
\FL
\begin{equation}
 \biggl(\frac{\partial a_t E_0(r)}
{\partial \beta_s}\biggr)_{\xi=1}=\frac {1}{2} V(r)\biggl(
\frac{\partial a}{\partial \beta} + \frac{a}{\beta} \biggr)
+\frac {1}{2}\biggl(\frac{\partial f(\beta)}{\partial \beta}
+\frac{f(\beta)}{\beta}\biggr).
\label{e5.34}
\end {equation}
 From Eqs.~(\ref{e5.27a}), (\ref{e5.27b}), (\ref{e5.28a}),
(\ref{e5.28b}) and (\ref{e5.33}), (\ref{e5.34}) one can get
the energy sum rules in Eqs.~(\ref{e5.1}) and~(\ref{e5.2}).

   Now we proceed to consider the generalization of these
sum rules to the finite temperature case.
In the limit of finite temperature ($T$) and infinite volume
the potential $V(r)$ would be a function of $r$ and $T$
(or $\beta$) because of $T=1/N_t a(\beta)$. For example,
in the confined phase $V(r)\sim \kappa r$, with the string
tension $\kappa=\kappa(\beta)$ for fixed lattice size $N_t$,
as discussed in Sec.~\ref{subsec: stc}. In this case one
can write, $V(r)=V(r,\beta)$. Then Eq.~(\ref{e5.31})
should be rewritten as,
\begin{equation}
 \biggl(\frac{\partial a_t V(r,\beta)}
{\partial \beta_t}\biggr)_{\xi=1}= \frac {1}{2}\biggl(
\frac{\partial a V(r,\beta)}{\partial \beta}
-\frac{a}{\beta}V(r,\beta) \biggr).
\label{e5.35}
\end {equation}
Similarly one has
\begin{equation}
 \biggl(\frac{\partial a_t V(r,\beta)}
{\partial \beta_s}\biggr)_{\xi=1}= \frac {1}{2}\biggl(
\frac{\partial a V(r,\beta)}{\partial \beta}
+\frac{a}{\beta}V(r,\beta) \biggr).
\label{e5.36}
\end {equation}
The self-energy part $f(\beta)/a$ does not change in this case.
Therefore, at finite temperature Michael sum rules in
Eqs.~(\ref{e5.1}) and (\ref{e5.2}) should be modified as
following,
\begin{equation}
 E_{\rm el}(r)
=-\frac{1}{2}\biggl(\frac{\beta}{a}\frac{\partial [aV(r)]}
{\partial \beta}-V(r)\biggr)
-\frac{1}{2a}\biggl(\beta \frac{\partial f}{\partial \beta}-f\biggr),
\label{e5.37}
\end {equation}
\begin{equation}
 E_{\rm ma}(r)
=\frac{1}{2}\biggl(\frac{\beta}{a}\frac{\partial [aV(r)]}
{\partial \beta}+V(r)\biggr)
+\frac{1}{2a}\biggl(\beta \frac{\partial f}{\partial \beta}+f\biggr).
\label{e5.38}
\end {equation}
 These modified sum rules can be obtained from the original ones
in Eqs.~(\ref{e5.1}) and (\ref{e5.2}) by the replacement,
\begin{equation}
 V(r)\frac{\partial a}{\partial \beta}\rightarrow
\frac{\partial [aV(r, \beta )]}{\partial \beta}.
\label{e5.39}
\end {equation}

   We expect that the modified sum rules would account for
the finite temperature effects on the $q\bar q$ system.
Although Eqs.~(\ref{e5.37}) and~(\ref{e5.38}) are
derived by considering the Wilson loop representation
of the $q\bar q$ pair, we expect that they can be applied
in describing the flux distributions $f'_{\mu \nu}$
in Eq.~(\ref{e3.5}), which involves the Polyakov loops,
because both the Wilson loop, $W$, and Polyakov loops,
$P(0)P^{\dagger}(r)$, represent a $q\bar q$ pair.

\subsection{Analysis of Flux Data }
\label{subsec: afd}

   We calculated the flux distributions $f'_{\mu \nu}$
of Eq.~(\ref{e3.5}) on various lattices, which cover the
regions of both confined and unconfined phases.
The detailed analysis of the $q\bar q$ flux distributions is
presented elsewhere~\cite{ph}. In this part we shall
compare our flux data with the predictions of modified
Michael sum rules in Eqs.~(\ref{e5.37}) and (\ref{e5.38}).

  As we mentioned in Sec.~\ref{subsec: mfd}, we used the
multihit technique~\cite{wh} in the flux measurements.
The flux data is only good when the {\it test charge}
(plaquette) is away from the $q\bar q$ sources
(Polyakov loops). Thus we only consider the flux data on the
middle transverse slice between the $q\bar q$ pair at large
separations (i.e., $r\ge 3a$).

  In the confined phase string formation is expected to
occur. When the separation, $r$, of a $q\bar q$ pair is
large enough, the flux energy stored in the flux tube per unit
length between the $q\bar q$ should equal to the string
tension $\kappa$. In this case one can obtain the color
electric and magnetic energy of the flux tube per unit length
from Eqs.~(\ref{e5.37}) and (\ref{e5.38}),
\begin{equation}
 \sigma_{\rm el}=\frac{dE_{el}(r)}{dr}
=-\frac{1}{2}\biggl(\frac{\beta}{a}\frac{\partial [a\kappa]}
{\partial \beta}-\kappa\biggr),
\label{e5.40}
\end {equation}
\begin{equation}
\sigma_{\rm ma}=\frac{dE_{ma}(r)}{dr}
=\frac{1}{2}\biggl(\frac{\beta}{a}\frac{\partial [a\kappa]}
{\partial \beta}+\kappa\biggr).
\label{e5.41}
\end {equation}
The corresponding flux action is
\begin{equation}
 \sigma_A=\sigma_{\rm el}-\sigma_{\rm ma}
=-\frac{\beta}{a}\frac{\partial [a\kappa]}{\partial \beta},
\label{e5.42}
\end {equation}
which is the integration of the action density on the transverse
plane. And the corresponding flux energy is
\begin{equation}
 \sigma_E=\sigma_{\rm el}+\sigma_{\rm ma}
=\frac{dV(r)}{dr}=\kappa(\beta),
\label{e5.43}
\end {equation}
which agrees with our expectation for string formation.

   Eq.~(\ref{e5.42}) can also be obtained from the action sum
rule of Eq.~(\ref{e5.17}). The R.H.S. of Eq.~(\ref{e5.42}) is,
 $\frac{\beta}{a}\frac{\partial (a\kappa)}{\partial \beta}
=\kappa \frac{\partial \ln a}{\partial \ln\beta}+
\beta \frac{\partial \kappa}{\partial \beta}$.
The first term can be estimated from the scaling relation
of Eq.~(\ref{e4.2}), e.g.,
$\frac{\partial \ln a}{\partial \ln\beta}\sim -8$ for $\beta\sim 2.4$,
which is consistent with the flux measurement results of
Ref.~\cite{hw}. The second term
$\beta \frac{\partial \kappa}{\partial \beta}$ can be
estimated from the fitting function in Eq.~(\ref{e4.8}),
and it is negative because $\kappa$ decreases with $\beta$
for fixed lattice size $N_t$, as shown in Eq.~(\ref{e4.5}).
So Eq.~(\ref{e5.42}) implies that $\sigma_A \gg \sigma_E=\kappa$
for cases we consider. Also, Eqs.~(\ref{e5.40}) and~(\ref{e5.41})
show that the electric part of the flux energy $\sigma_{\rm el}$
is positive, but the magnetic part $\sigma_{\rm ma}$ negative,
their magnitudes are of the same order with $\sigma_{\rm el}$
slightly larger. These predictions can be summarized as below,
\begin{eqnarray}
\sigma_{\rm el}&&>0 \qquad \mbox{and} \qquad \sigma_{\rm ma}<0,
\nonumber \\
\sigma_E &&=\sigma_{\rm el}+\sigma_{\rm ma}=\kappa >0, \nonumber \\
\sigma_A && \gg \sigma_E;
\label{e5.44}
\end{eqnarray}
where these flux energies (action) are defined in
the confined phase.

  From our flux data we can calculate the values of
$\sigma_{\rm el}$, $\sigma_{\rm ma}$, $\sigma_{E}$ and
$\sigma_{A}$, which are the energies (action) stored in the
center slice per unit length between the $q\bar q$ pair.
We find that our data agree with the predictions of
Eq.~(\ref{e5.44}) qualitatively. However, it is difficult
to measure $\sigma_E$ accurately because of the strong
cancellation between two terms, $\sigma_{\rm el}$ and
$\sigma_{\rm ma}$, which have opposite signs. This
problem was also discussed by Refs.~\cite{mis,hw}.

  In the following we proceed to study the prediction
of the action sum rule in Eq.~(\ref{e5.42}). As we
mentioned above, the R.H.S. of Eq.~(\ref{e5.42}) contains
two terms, the first term,
$\kappa \frac{\partial \ln a}{\partial \ln \beta}$,
was predicted by the original Michael sum rules in
Eqs.~(\ref{e5.1}) and (\ref{e5.2}), the second one,
$\beta \frac{\partial \kappa}{\partial \beta}$, is a
new term, which is only predicted by the modified sum
rules of Eqs.~(\ref{e5.37}) and (\ref{e5.38}).
We expect that the second term describes finite
temperature effects.

 From the scaling relation $a(\beta)$ of Eq.~(\ref{e4.2})
one can estimate the quantity,
\begin{equation}
 -\frac{\partial \ln a}{\partial \ln\beta}
= -\frac{51}{121}+\frac{3\pi^2}{11}\beta -
\frac{d_2+2d_3\beta}{\Lambda_L^{-1}(\beta)}\beta,
\label{e5.45}
\end {equation}
with $\Lambda_L^{-1}(\beta)$ defined in Eq.~(\ref{e4.2}).
And by using the fitting function $\kappa(\beta)$ in
Eq.~(\ref{e4.8}) one can calculate the derivative,
\begin{equation}
 \frac{\partial \kappa}{\partial \beta}
= -\frac{\kappa_0}{\beta_c}\delta
(1-\frac{\beta}{\beta_c})^{\delta -1},
\label{e5.46}
\end {equation}
with the constants $\kappa_0$ and $\delta$
given by Eqs.~(\ref{e4.1}) and (\ref{e4.9}).

  By similar considerations that lead to Eq.~(\ref{e5.42}),
one can also obtain the prediction about $\sigma_A$
from the original Michael sum rules in Eqs.~(\ref{e5.1})
and (\ref{e5.2}). The result is
\begin{equation}
 (\sigma_A)_0
=-\kappa \frac{\partial \ln a}{\partial \ln\beta},
\label{e5.47}
\end {equation}
which is just the first term of the R.H.S. in Eq.~(\ref{e5.42}).

   In Table~\ref{t5.2} we list the predictions of
Eq.~(\ref{e5.42}) obtained by substituting Eqs.~(\ref{e5.45}) and
(\ref{e5.46}), and the measured $\sigma_A$ data in the
confined phase. Since in the confined phase
the value of $\sigma_A$ should not change with the $q\bar q$
separation $r$ in the limit of $r\rightarrow \infty$,
we then choose the $\sigma_A$ data at a moderate value of $r$
as the asymptotic value of $\sigma_A$ ($r\rightarrow \infty$),
because error bars are large at very large $r$.
In Table~\ref{t5.2} most $\sigma_A$ data were calculated
from the flux measurements of $q\bar q$ pair at
$r=4 a$, in few cases we choose
the data at $r=3a$ or $5a$, depending on the quality of data.

   In Fig.~\ref{f5.2} we plot the predictions of Eqs.~(\ref{e5.42})
and (\ref{e5.47}) respectively, and compare them
with the measured $\sigma_A$ data on lattices of $N_t=4$.
{}From this figure one can see that the measured data
are consistent with the prediction of Eq.~(\ref{e5.42}),
but have large difference from the prediction of
Eq.~(\ref{e5.47}). Especially, in the transition region
($\beta\approx \beta_c$) Eq.~(\ref{e5.47}) predicts that
$\sigma_A$ approaches zero as $\beta\rightarrow \beta_c$.
However, our $\sigma_A$ data have large values in this
region, which agree with the prediction of Eq.~(\ref{e5.42}).
As the temperature decreases ($T\rightarrow 0$
or $\beta\rightarrow 0$), both predictions coincide.
However, it is difficult to obtain data of clear signal
in the small $\beta$ region (i.e., $\beta < 2.25$).

  We also notice that in Fig.~\ref{f5.2} the $\sigma_A$ data
in the confined region ($\beta < \beta_c$) have some
discrepancies from the prediction of Eq.~(\ref{e5.42}).
This may be due to finite-size effects and finite lattice
spacing effects of lattices, and to the large fluctuations
in the confined phase because confinement corresponds to disorder.

  To compare the behaviors of $\sigma_A$ in both phases,
in Fig.~\ref{f5.2} we also plot the $\sigma_A$ data in the
unconfined region (i.e., $\beta > \beta_c$). These data were
measured at large $q\bar q$ separations, i.e., $r=6a$,
which still have clear signals, as shown in Table~\ref{t5.3}.
Since in the unconfined phase there is no string formation,
one expects that $\sigma_A$ vanishes at large $r$.
{}From this figure one can see that near the transition point,
i.e., $\beta \sim 2.30$, the $\sigma_A$ data in the
unconfined region decrease rapidly with $\beta$,
and becomes very small beyond the transition region.
This agrees with the expectation. The fact that our $\sigma_A$
data in the unconfined region do not vanish may be due to
the following factors, in the transition region
finite-size effects are large, beyond this region
the contribution from the self-energy of the $q\bar q$ pair
still exists, because the $q\bar q$ separation, $r=6a$,
is not large enough to approach the asymptotic region.

\bigskip
\section{Summary}
\label{sec: sum}

   We studied the SU(2) finite temperature phase transition
by Monte Carlo simulations.
To transform the measurements from lattice units to physical
units the lattice asymptotic scaling relation $a(\beta)$
is extrapolated into the non-perturbative region.
The behavior of string tension with temperature is also
studied, We find that the string tension
data agree very well with the fitting function,
${\kappa(T)=\kappa_0(1-\frac{T}{T_c})^\alpha}$ for $T<T_c$.

   We also measured the flux distribution of the $q\bar q$ pair
at various temperatures. To check Michael sum rules with the
flux data, a complete derivation of the sum rules are presented,
and a generalization of the sum rules is suggested to
account for the finite temperature effects.
We found that our flux data are consistent with the prediction
of the generalized sum rules. Our data shows explicitly that
the $q\bar q$ flux distribution has different behaviors in
the two phases. In the confined phase the asymptotic value of
the center slice action, $\sigma_A$ ($r\rightarrow \infty$),
has large values. However, across the transition
region, $\sigma_A$ becomes very small in the unconfined
phase. This agrees with our expectation that
 string formation occurs in the confined phase,
but disappears in the unconfined phase.

\bigskip
\nonum
\section{Acknowledgments}

  We wish to thank C. Michael, B.A. Berg, J. Wosiek,
D.A. Browne and V. Singh for many fruitful discussions on
this problem. This research was supported by the U.S. Department of
Energy under Grant No. DE-FG05-91ER40617.

\newpage
\narrowtext
\appendix{ }

\bigskip

  To prove Eqs.~(\ref{e5.29a}) and (\ref{e5.29b}), we notice
that Eqs.~(\ref{e5.23a}) and (\ref{e5.23b}) implies
that $\beta_s$, $\beta_t$ and $\beta$, $\xi$ are two
equivalent sets of variables, that is,
${F(\beta_s, \beta_t)=F(\beta_s(\beta,\xi),\beta_t(\beta,\xi))}$.
Then we take the partial derivatives of $F$
with respect to $\beta$ and $\xi$ respectively, one has
\begin{equation}
 \biggl(\frac {\partial F}{\partial \beta} \biggr)_{\xi}
= \biggl(\frac {\partial F}{\partial \beta_t}\biggr)_{\beta_s}
 \biggl(\frac {\partial \beta_t}{\partial \beta}\biggr)_{\xi}
+\biggl(\frac {\partial F}{\partial \beta_s}\biggr)_{\beta_t}
 \biggl(\frac {\partial \beta_s}{\partial \beta}\biggr)_{\xi},
\label{ea1}
\end {equation}
\begin{equation}
 \biggl(\frac {\partial F}{\partial \xi} \biggr)_{\beta}
= \biggl(\frac {\partial F}{\partial \beta_t}\biggr)_{\beta_s}
 \biggl(\frac {\partial \beta_t}{\partial \xi}\biggr)_{\beta}
+\biggl(\frac {\partial F}{\partial \beta_s}\biggr)_{\beta_t}
 \biggl(\frac {\partial \beta_s}{\partial \xi}\biggr)_{\beta}.
\label{ea2}
\end {equation}

  From Eqs.~(\ref{e5.23a}) and~(\ref{e5.23b}) one can get that,
when $\beta$ is large (i.e., $\beta \rightarrow \infty$),
\begin{eqnarray}
 \biggl(\frac {\partial \beta_t}
{\partial \beta}\biggr)_{\xi=1} &&=\biggl(\frac {\partial \beta_s}
{\partial \beta}\biggr)_{\xi=1}=1; \nonumber \\
 \biggl(\frac {\partial \beta_s}{\partial \xi}
\biggr)_{\xi=1} &&=-\beta+2N c'_s |_{\xi=1}, \nonumber \\
 \biggl(\frac {\partial \beta_t}{\partial \xi}
\biggr)_{\xi=1} &&=\beta+2N c'_t |_{\xi=1}.
\label{ea3}
\end {eqnarray}
where $N$ denotes for SU($N$), and $c'$ the $\xi$-derivative of $c$.
Substituting Eq.~(\ref{ea3}) into Eqs.~(\ref{ea1}) and
(\ref{ea2}) yields
\begin{equation}
  \biggl(\frac {\partial F}{\partial \beta}
\biggr)_{\xi=1}= \biggl(\frac {\partial F}{\partial \beta_t}
 \biggr)_{\xi=1}+\biggl(\frac {\partial F}{\partial \beta_s}
 \biggr)_{\xi=1},
\label{ea4}
\end {equation}
\begin{equation}
 \biggl(\frac {\partial F}{\partial \xi}
\biggr)_{\xi=1}= \biggl [\frac {\partial F}{\partial \beta_t}
  (\beta +2N c'_t) +\frac {\partial F}{\partial \beta_s}
  (-\beta+2N c'_s)\biggr ]_{\xi=1}.
\label{ea5}
\end {equation}

F. Karsch has studied the coefficients $c_s$ and $c_t$~\cite{ka}.
His results show that the derivatives $c'_s|_{\xi=1}$, $c'_t|_{\xi=1}$
vanish for sufficient large lattice coupling constant, $\beta$.
For example, as $\beta \ge 2.2$, the values of
$c'_s|_{\xi=1}$ and $c'_t|_{\xi=1}$ of SU(2) LGT are about $0.1$,
which is much less than the values of $\beta$. For large $\beta$
cases Eqs.~(\ref{ea4}) and (\ref{ea5}) become
\begin{equation}
 \biggl(\frac {\partial F}{\partial \beta}
\biggr)_{\xi=1}= \biggl(\frac {\partial F}{\partial \beta_t}
 \biggr)_{\xi=1}+\biggl(\frac {\partial F}{\partial \beta_s}
 \biggr)_{\xi=1} ,
\label{ea6}
\end {equation}
\begin{equation}
 \biggl(\frac {\partial F}{\partial \xi}
\biggr)_{\xi=1}= \beta \biggl(\frac {\partial F}{\partial \beta_t}
 \biggr)_{\xi=1} -\beta \biggl(\frac {\partial F}
{\partial \beta_s}\biggr)_{\xi=1} .
\label{ea7}
\end {equation}
  After carrying out some simple algebra, we can obtain
Eqs.~(\ref{e5.29a}) and (\ref{e5.29b}).


\widetext

\begin{table}
\caption{ The correspondence of the lattice spacing $a$ and the
coupling constant $\beta$ for SU(2) LGT, extracted from
the string tension data of Refs.~\cite{so,phm}. }
\begin{tabular}{ccc}
$\beta$ &  $a(\beta)$ (fm)  & $\Lambda_L^{-1}$ (fm)  \\
\tableline
  2.22  & 0.1981 (61)   & 27.41 (84)   \\
  2.30  & 0.1616 (47)   & 27.32 (79)   \\
  2.40  & 0.1210 (5)    & 26.30 (11)   \\
  2.50  & 0.0843 (8)    & 23.57 (22)   \\
\end{tabular}
\label{t4.1}
\end{table}
\bigskip

\begin{table}
\caption{ The raw string tension data $\kappa$ measured in
lattice units on lattices of the size,
$4\times 9^2\times 65$, $4\times 11^2\times 65$,
$6\times 7^2\times 65$ and $6\times 11^2\times 37$. }
\begin{tabular}{cccccc}
 $N_t=4$  & $4\times 9^2\times 65$    & $4\times 11^2\times 65$
& $N_t=6$ & $6\times 7^2\times 65$    & $6\times 11^2\times 37$    \\
\tableline
  $\beta$ & $\sqrt{\kappa} a$  & $\sqrt{\kappa} a$
& $\beta$ & $\sqrt{\kappa} a$  & $\sqrt{\kappa} a$  \\
\tableline
  2.25 & $0.301$ ($38$) & $0.313$ ($32$)
& 2.30 & $0.310$ ($48$)  &  0.377 (27)         \\
  2.28 & $0.242$ ($33$)  & $0.243$ ($29$)
& 2.36 & $0.248$ ($21$) &  0.284 (13)         \\
  2.29 & $0.224$ ($17$)  &  0.181 (33)
& 2.40 & $0.172$ ($30$)  & $0.217$ ($17$)     \\
  2.30 & $0.216$ ($19$)  & $0.203$ ($17$)
& 2.42 & $0.200$ ($17$)  &  0.177 (41)         \\
\end{tabular}
\label{t3.2}
\end{table}
\bigskip

\begin{table}
\caption{The string tension data $\kappa$ measured in
physical units on lattices of the size, $4\times 9^2\times 65$,
$4\times 11^2\times 65$, $6\times 7^2\times 65$ and
$6\times 11^2\times 37$,
which were calculated from the data in Table~\ref{t3.2}
by using the scaling relation $a(\beta)$ in Eq.(~\ref{e4.2}).}
\begin{tabular}{ccccc}
\multicolumn{3}{c}{$N_t=4$} & $4\times 9^2\times 65$  &
 $4\times 11^2\times 65$ \\
\tableline
$\beta$ & $T\Lambda_L^{-1}$ & $T$ (1/fm)
& $\kappa$ (GeV/fm)  & $\kappa$ (GeV/fm) \\
  \tableline
  2.25 & 37.29 & 1.271 & 0.46 (12) & 0.50 (10) \\
  2.28 & 40.20 & 1.401 & 0.36 (10) & 0.37 (9)  \\
  2.29 & 41.22 & 1.447 & 0.33 (5) & 0.22 (8)  \\
  2.30 & 42.27 & 1.494 & 0.33 (6) & 0.29 (5)  \\
\tableline
\multicolumn{3}{c}{$N_t=6$} &  $6\times 7^2\times 65$  &
 $6\times 11^2\times 37$ \\
\tableline
$\beta$ & $T\Lambda_L^{-1}$ & $T$ (1/fm)
& $\kappa$ (GeV/fm)  & $\kappa$ (GeV/fm) \\
\tableline
  2.30 & 28.18 & 0.996 & 0.68 (21) &  1.00 (14) \\
  2.36 & 32.76 & 1.213 & 0.64 (11) &  0.84 (8) \\
  2.40 & 36.23 & 1.386 & 0.40 (14) &  0.64 (10)  \\
  2.42 & 38.10 & 1.482 & 0.62 (11) &  0.49 (23)  \\
\end{tabular}
\label{t4.2}
\end{table}
\bigskip

\begin{table}
\caption{The predictions from Eq.~(\ref{e5.42}) and
the data of center slice action $\sigma_A$,
which were measured on lattices
$4\times 9^2\times 65$, $4\times 11^2\times 65$,
$6\times 7^2\times 65$ and $6\times 11^2\times 37$.
The quantities $\kappa (\beta)$,
$\beta \frac{\partial \kappa}{\partial \beta}$,
$\kappa\frac{\partial lna}{\partial ln\beta}$ and $\sigma_A$
have the physical unit GeV/fm. The values of
$\kappa (\beta)$ were estimated from Eq.~(\ref{e4.8}).
Near the transition point (i.e., $\beta_c\sim 2.30$ for $N_t=4$)
no stable prediction is obtained. }
\begin{tabular}{ccccccc}
\multicolumn{7}{c}{$N_t=4$}  \\
\tableline
$\beta$ & $\kappa (\beta)$
& -$\kappa \frac{\partial lna}{\partial ln\beta}$
& -$\beta \frac{\partial \kappa}{\partial \beta}$
&-$\kappa \frac{\partial lna}{\partial ln\beta}$
  -$\beta \frac{\partial \kappa}{\partial \beta}$
&  $(\sigma_A)_{4\cdot 9^2\cdot 65}$
&  $(\sigma_A)_{4\cdot 11^2\cdot 65}$  \\
\tableline
2.25 & 0.42 & 3.03 (47) & 4 (1) & 7 (2)  &  12.18 (23)
  & 8.70 (22)    \\
2.28 & 0.34 & 2.50 (48) & 9 (3) & 12 (3) &  10.95 (14)
  & 9.97 (13)   \\
2.29 & 0.29 & 2.12 (47) & 17 (6) & 19 (7) &  10.44 (12)
  & 9.39 (10)   \\
\end{tabular}
\label{t5.2}
\end{table}

\begin{table}
\caption{The $\sigma_A$ data in the unconfined phase
($\beta < \beta_c$), which were measured on lattices,
 $4\times 9^2\times 65$ and $4\times 11^2\times 65$.
The data are in the physical unit Gev/fm.}
\begin{tabular}{cccccc}
\multicolumn{5}{c}{$N_t=4$}  \\
\tableline
$\beta$   & 2.30  &  2.32     &  2.34     &  2.36   &    2.40   \\
$(\sigma_A)_{4\times 9^2\times 65}$ & 11.77 (11) & 10.40 (9) &
                   &  1.78 (4)  &  1.28 (4)  \\
$(\sigma_A)_{4\times 11^2\times 65}$ & 11.03  &      &  3.81 (4)
                   &  1.77 (3)  &  1.32 (4)  \\
\end{tabular}
\label{t5.3}
\end{table}
\bigskip


\figure{ Monte Carlo data for $\langle |P|\rangle$ vs. $\beta$
with the standard Wilson action, calculated from lattices
$4\times 9^2\times 65$ (circles), $4\times 11^2\times 65$
(squares), $6\times 7^2\times 65$ (triangles)
and $6\times 11^2\times 37$ (diamonds).
\label{f3.1} }

\figure{ Monte Carlo data of $\langle P(0)P(r)\rangle$ vs. $r/a$
in the confined phase, calculated on the lattice,
$4\times 11^2\times 65$ with $\beta =2.25$ (squares),
$\beta =2.28$ (triangles) and $\beta =2.29$ (diamonds).
These data are all in the confined region, with
$\beta < \beta_c$ ($\beta_c \sim 2.30$ for $N_t=4$).
The plot is in a frame with logarithmic $y$ axis.
\label{f3.2} }

\figure{ Monte Carlo data of $\langle P(0)P(r)\rangle$ vs. $r/a$
in the unconfined phase, calculated on the lattice
$4\times 9^2\times 65$ with $\beta =2.36$ (squares), and
$\beta =2.40$ (triangles).
\label{f3.3} }

\figure{ The plot of $\kappa$ vs. $T$ (1/fm) near
the transition point $T_c \sim 1.487$ (1/fm).
Data were calculated on lattices, $4\times 9^2\times 65$ (circles),
$4\times 11^2\times 65$ (squares) and $6\times 7^2\times 65$
(triangles). The solid line is the fitting function in
Eq.~(\ref{e4.6}) with $\alpha=0.35\pm 0.04$.
 The string tension $\kappa$ is in the physical unit GeV/fm.
\label{f4.1} }

\figure{ $\kappa$ vs. $\beta $ for $N_t=4$, the transition
point is chosen as $\beta_c= 2.2985$.
Data were calculated on lattices, $4\times 9^2\times 65$ (squares),
$4\times 11^2\times 65$ (triangles).
The solid line is the fitting function in Eq.~(\ref{e4.8})
with $\delta=0.22\pm 0.03$. The string tension $\kappa$ is in the
physical unit GeV/fm.
\label{f4.2} }

\figure{The Wilson loop $W$ of the temporal size $na$ and the
spatial size $r$. The plaquette $P_1$ is outside the Wilson loop $W$,
and the plaquette $P_2$ is inside the Wilson loop $W$.
\label{f5.1} }

\figure{ The plot of the predictions of $\sigma_A$ vs.
$\beta$ from Eq.~(\ref{e5.42}) (solid lines), and Eq.~(\ref{e5.47})
(dashed lines) in the confined region ($\beta < \beta_c$).
The two solid lines represent the upper and
lower limits predicted by Eq.~(\ref{e5.42}).
The data were measured on lattices, $4\times 9^2\times 65$
(squares) and $4\times 11^2\times 65$ (triangles). For comparison
the data in the unconfined region ($\beta > \beta_c$) are also
shown, which were measured on the same lattices,
$4\times 9^2\times 65$ (circles) and
$4\times 11^2\times 65$ (diamonds).
Here $\sigma_A$ has the physical unit GeV/fm, and the
transition point $\beta_c$ is indicated by the up arrow.
\label{f5.2} }

\end{document}